\documentclass[12pt]{article}

\pdfoutput=1
\usepackage{jheppub}
\usepackage{graphicx,epsfig,amsmath,amssymb,mathtools}
\usepackage{subcaption}
\usepackage{xspace}
\graphicspath{{plots/}}

\def\be{\begin{equation}}
\def\ee{\end{equation}}
\def\bea{\begin{eqnarray}}
\def\eea{\end{eqnarray}}

\newcommand{\powheg}{{\tt POWHEG}}
\newcommand{\powhegbox}{{\tt POWHEG-BOX}}

\newcommand{\gosam}{\textsc{GoSam}{}}

\newcommand{\ninja}{{\tt Ninja}}
\newcommand{\avholo}{{\tt OneLOop}}
\newcommand{\pythia}{{\tt Pythia\,8.2}\xspace}
\newcommand{\pyth}{{\tt Pythia}\xspace}

\newcommand{\herwig}{{\tt Herwig\,7.1}\xspace}
\newcommand{\herw}{{\tt Herwig}\xspace}
\newcommand{\hdamp}{{\tt hdamp}}

\newcommand{\GeV}{\ensuremath{\mathrm{GeV}}}
\newcommand{\TeV}{\ensuremath{\mathrm{TeV}}}

\newcommand{\chhh}{\kappa_{\lambda}}
\newcommand{\mhh}{m_{\mathrm{hh}}}
\newcommand{\pth}{p_T^{\mathrm{h}}}
\newcommand{\pthh}{p_T^{\mathrm{hh}}}
\newcommand{\ftapprox}{FT$_{\mathrm{approx}}$}

\title{Probing the trilinear Higgs boson coupling in di-Higgs production at NLO QCD including parton shower effects}

\author[a]{G.~Heinrich,}
\author[b]{S.~P.~Jones,}
\author[c]{M.~Kerner,}
\author[a]{G.~Luisoni,}
\author[a]{L.~Scyboz}

\affiliation[a]{Max Planck Institute for Physics, F\"ohringer Ring 6,  80805 M\"unchen, Germany}
\affiliation[b]{Theoretical Physics Department, CERN, Geneva, Switzerland}
\affiliation[c]{Physik-Institut, Universit{\"a}t Z{\"u}rich, Winterthurerstrasse 190, 8057 Z{\"u}rich, Switzerland}

\emailAdd{gudrun@mpp.mpg.de}
\emailAdd{s.jones@cern.ch}
\emailAdd{mkerner@physik.uzh.ch}
\emailAdd{luisonig@gmail.com}
\emailAdd{scyboz@mpp.mpg.de}

\preprint{{\small  CERN-TH-2019-028, MPP-2019-50, ZU-TH 09/19}}

\abstract{
 We present results for Higgs boson pair 
  production with variations of the trilinear Higgs boson self coupling at next-to-leading order (NLO) in QCD including the full top quark mass dependence.
  Differential results for the LHC at 14\,TeV are presented, and we discuss
  the implications of anomalous trilinear couplings as well as differences between the \pyth and \herw parton showers in combination with \powheg.
 The implementation of the NLO QCD calculation with variable Higgs boson self coupling is made publicly available in the {\tt POWHEG-BOX-V2}
  Monte Carlo framework. A simple method for using the new implementation to study also variations of the top quark Yukawa coupling is described.
}

\begin{document}

\maketitle

\section{Introduction}

The Higgs potential is currently the least explored part of the Standard Model (SM), measurements of the Higgs boson self-coupling(s) may therefore offer surprises.
Although the Higgs boson couplings to vector bosons and third generation fermions are increasingly well measured~\cite{Khachatryan:2016vau,Aaboud:2017vzb,ATLAS:2018doi,Sirunyan:2018koj,Sirunyan:2018sgc}, constraints on the trilinear coupling $\lambda$ are relatively weak due to the small Higgs boson pair production cross sections~\cite{Baglio:2012np,Frederix:2014hta}.  
Nonetheless, measurements of double Higgs production in gluon fusion, combining various decay channels,  have led to impressive experimental results already~\cite{Sirunyan:2018two,Sirunyan:2018iwt,Aad:2019uzh,Aaboud:2018ftw}, 
the most stringent constraints on the trilinear coupling being $-5\leq \kappa_\lambda\leq 12.1$
 at 95\% confidence level~\cite{Aad:2019uzh}, based on the assumption that all other couplings have SM values.
Individual limits on $\kappa_\lambda$ based on EFT benchmarks representing a certain combination of BSM couplings which leads to characteristic kinematic distributions~\cite{Carvalho:2015ttv,Carvalho:2016rys,Buchalla:2018yce} have also been extracted~\cite{Sirunyan:2018two,Sirunyan:2018iwt}.
Therefore, the determination of the trilinear coupling has entered a level of precision where the assumption that the full NLO QCD corrections do not vary much with $\kappa_\lambda$, which has been used in the experimental analysis so far, needs to be revised.
The variations of the K-factors with $\kappa_\lambda$ are mild in the $m_t\to \infty$ limit, where NLO~\cite{Grober:2015cwa,Grober:2017gut} and NNLO~\cite{deFlorian:2017qfk} corrections have been calculated within an effective Lagrangian framework.
However, it will be shown in this paper that the NLO K-factor varies by about 35\% as $\kappa_\lambda$ is varied between $-1$ and 5 once the full top quark mass dependence is taken into account. 

The question of how large or small $\chhh$ can be from a theory point of view is not easy to answer in a model independent way. 
Recent work based on rather general concepts like vacuum stability and perturbative unitarity suggests that $|\chhh|\lesssim 4$ for a New Physics scale in the few TeV range~\cite{Falkowski:2019tft,Chang:2019vez,DiLuzio:2017tfn,DiVita:2017eyz}.
More specific models can lead to more stringent bounds, see e.g. Refs.~\cite{Braathen:2019pxr,Basler:2018dac,Babu:2018uik,Lewis:2017dme}.
Recent phenomenological studies about the precision that could be reached for the trilinear coupling at the (HL-)LHC and future hadron colliders can be found for example in Refs.~\cite{Cepeda:2019klc,Li:2019uyy,Homiller:2018dgu,Bizon:2018syu,Goncalves:2018yva,Kim:2018uty,Adhikary:2017jtu,Corbett:2017ieo,Alves:2017ued,Cao:2016zob,Azatov:2015oxa}.

\medskip

Higgs boson pair production in gluon fusion in the SM has been calculated at leading order in Refs.~\cite{Eboli:1987dy,Glover:1987nx,Plehn:1996wb}.
The NLO QCD corrections with full top quark mass dependence became available more recently~\cite{Borowka:2016ehy,Borowka:2016ypz,Baglio:2018lrj}.
The NLO results of Refs.~\cite{Borowka:2016ehy,Borowka:2016ypz} have been supplemented by soft-gluon resummation at small transverse momenta of the Higgs boson pair~\cite{Ferrera:2016prr}
and by parton shower effects~\cite{Heinrich:2017kxx,Jones:2017giv}.
Before the full NLO QCD corrections became available, the $m_t\to\infty$ limit, sometimes also called Higgs Effective Field Theory~(HEFT) approximation,
has been used in several forms of approximations. 
In this limit, the NLO corrections were first calculated in 
Ref.~\cite{Dawson:1998py} using the so-called ``Born-improved HEFT'' approximation, 
which involves rescaling the NLO results in the $m_t\to\infty$ limit by a factor $B_{\rm FT}/B_{\rm HEFT}$, where $B_{\rm FT}$
denotes the LO matrix element squared in the full theory.
In Ref.~\cite{Maltoni:2014eza} an approximation called
``\ftapprox'', was introduced, which contains the real radiation matrix elements 
with full top quark mass dependence, while the virtual part is
calculated in the Born-improved HEFT approximation.

The NNLO QCD corrections in the $m_t\to\infty$ limit have been computed in Refs.~\cite{deFlorian:2013uza,deFlorian:2013jea,Grigo:2014jma,deFlorian:2016uhr}. 
These results have been improved in various ways: they have been supplemented by an expansion in $1/m_t^2$ in~\cite{Grigo:2015dia}, and soft gluon resummation has been performed at NNLO+NNLL level in~\cite{deFlorian:2015moa}. 
 The calculation of Ref.~\cite{deFlorian:2016uhr} has been combined with results including the top quark mass dependence as far as available in Ref.~\cite{Grazzini:2018bsd}, and the latter has been supplemented by soft gluon resummation in Ref.~\cite{deFlorian:2018tah}. 

The scale uncertainties at NLO are still at the 10\% level, while they are decreased to about 5\% when including the NNLO corrections.
The uncertainties due to the chosen top mass scheme have been assessed in Ref.~\cite{Baglio:2018lrj}, where the full NLO corrections, including the possibility to switch between pole mass and $\overline{\mathrm{MS}}$ mass, have been presented.

Analytic approximations for the top quark mass dependence of the two-loop amplitudes in the NLO calculation have been studied in Refs.~\cite{Grober:2017uho,Bonciani:2018omm,Xu:2018eos,Davies:2018ood} and complete analytic results in the high energy limit have been presented in Ref.~\cite{Davies:2018qvx}. The formalism of an expansion for large top quark mass has been applied recently to calculate partial real-radiation corrections to Higgs boson pair production at NNLO in QCD~\cite{Davies:2019xzc}.

%%%%%%%%%%%%%%%

In this work we study the dependence of total cross sections and differential distributions on the trilinear Higgs boson coupling, assuming that the BSM-induced deviations in the other couplings are at the (sub-)percent level.
The study is based on results at NLO QCD with full top quark mass dependence for Higgs boson pair production in gluon fusion described in Refs.~\cite{Borowka:2016ehy,Borowka:2016ypz}. 
While it is unlikely that New Physics alters just the Higgs boson self-couplings but leaves the Higgs couplings to vector bosons and fermions unchanged, it can be assumed that the deviations of the measured Higgs couplings from their SM values are so small that they have escaped detection at the current level of precision, for recent overviews see e.g. Refs.~\cite{Cepeda:2019klc,Dawson:2018dcd,Brooijmans:2018xbu,deFlorian:2016spz}.

Measuring Higgs boson pair production is a direct way to access the trilinear Higgs coupling. The trilinear and quartic couplings can also be constrained in an indirect way, through measurements of processes which are sensitive to the Higgs boson self-couplings via electroweak corrections~\cite{Gorbahn:2019lwq,Nakamura:2018bli,Borowka:2018pxx,Bizon:2018syu,Kilian:2018bhs,Vryonidou:2018eyv,Maltoni:2018ttu,Maltoni:2017ims,Kribs:2017znd,Degrassi:2017ucl,Bizon:2016wgr,Degrassi:2016wml,Gorbahn:2016uoy,McCullough:2013rea}.
Such processes offer important complementary information, however they are more susceptible to other BSM couplings entering the loop corrections 
at the same level, and therefore the limits on $\chhh$ extracted this way may be more model dependent than the ones extracted from the direct production of Higgs boson pairs.

For Higgs Boson pair production, due to the destructive interference in the squared amplitude between contributions containing $\lambda$ and those without the Higgs boson self-coupling (corresponding to triangle- and box diagrams, respectively, at LO), 
%where the destructive interference is maximal at $\lambda\sim 2.4$, 
small changes in $\lambda$ modify the interference pattern and can therefore have a substantial effect on the total cross section and differential distributions.

In order to obtain a fully-fledged NLO generator which also offers the possibility of parton showering, we have implemented our calculation in the 
\powhegbox~\cite{Nason:2004rx,Frixione:2007vw,Alioli:2010xd}, building on the SM code presented in Ref.~\cite{Heinrich:2017kxx}.

The dependence of the K-factors on the value of $\lambda$ (and other BSM couplings) is stronger than the $m_t\to\infty$ limit may suggest, as shown in Ref.~\cite{Buchalla:2018yce}. This is particularly true for differential K-factors. For example, in the boosted regime, which is sometimes used by the experiments when reconstructing the $H \rightarrow b\bar{b}$ decay channel, Higgs bosons with a large-$p_T$ are involved. At large-$p_T$ the top quark loops are resolved and the $m_t\to\infty$ limit is invalid.
The top quark mass corrections in the large $\mhh$ or $\pth$ regime are of the order of 20-30\% or higher, and increase with larger centre-of-mass energy (e.g. $\sqrt{s}=27$ (HE-LHC) or 100\,TeV (FCC-hh)), these corrections clearly exceed the scale uncertainties and therefore have to be taken into account.

The purpose of this paper is twofold: Based on our differential results, we discuss how the deviations from the SM, resulting from non-SM $\lambda$ values, can be identified based on the distributions for the Higgs boson pair invariant mass and Higgs boson transverse momentum distributions. 
In addition, we present the updated public code {\tt POWHEG-BOX-V2/ggHH}, where the user can choose the value of the trilinear coupling as an input parameter.
We also explain how variations of the top-Higgs Yukawa coupling can be studied using this code.
Further, we compare the fixed order results to results obtained by matching the NLO calculation to a parton shower. In particular, we compare results from the \pythia~\cite{Sjostrand:2014zea} and \herwig~\cite{Bellm:2017bvx} parton showers and assess the parton-shower related uncertainties.

This paper is organised as follows. In Section~\ref{sec:calculation} we briefly describe the calculation and give instructions for the usage of the program within the \powhegbox. Section~\ref{sec:results} contains the discussion of our results, focusing in the first part on variations of $\chhh$ and in the second part on differences between showered results. We present our conclusions in Section~\ref{sec:conclusions}.

\section{Overview of the calculation}
\label{sec:calculation}

The calculation builds on the one presented in Ref.~\cite{Heinrich:2017kxx} and therefore will be described only briefly here. 

The leading order amplitude in the full theory and  all the
amplitudes in the $m_t\to\infty$ limit were implemented analytically, whereas the
one-loop real radiation contribution and the two-loop virtual
amplitudes in the full SM rely on numerical or semi-numerical
codes.
The real radiation matrix elements in the full SM were implemented
using the interface~\cite{Luisoni:2013cuh}
between \gosam~\cite{Cullen:2011ac,Cullen:2014yla} and
the \powhegbox~\cite{Nason:2004rx,Frixione:2007vw,Alioli:2010xd}, modified
accordingly to compute the real corrections to the loop-induced Born amplitude. 
%The one-loop real amplitudes we generated with the version 2.0 of \gosam{}~\cite{Cullen:2014yla}, 
%which uses {\tt Qgraf}~\cite{Nogueira:1991ex}, \form~\cite{Kuipers:2012rf} and
%{\tt spinney}~\cite{Cullen:2010jv} for the generation of the Feynman
%diagrams, and offers a choice from {\tt Samurai}~\cite{Mastrolia:2010nb,vanDeurzen:2013pja}, {\tt golem95C}~\cite{Binoth:2008uq,Cullen:2011kv,Guillet:2013msa}
%and \ninja{}~\cite{Peraro:2014cba} for the
%reduction. 
 At run time the amplitudes were computed using
\ninja{}~\cite{Peraro:2014cba},  {\tt golem95C}~\cite{Binoth:2008uq,Cullen:2011kv} and \avholo{}~\cite{vanHameren:2010cp}
for the evaluation of the scalar one-loop integrals.
The stability of the amplitudes in the collinear limits has been improved by a better detection of instabilities in the real radiation 
and the use of the scalar four-point function from {\tt VBFNLO}~\cite{Arnold:2008rz,Baglio:2014uba}.

For the virtual corrections, containing two-loop amplitudes, we have used the results of the
calculation presented in Refs.~\cite{Borowka:2016ehy,Borowka:2016ypz},
which used also {\sc Reduze}\,2~\cite{vonManteuffel:2012np} and {\sc
 SecDec}\,3~\cite{Borowka:2015mxa}.

The values for the Higgs boson and top quark masses have been set to
$m_h=125$\,GeV and $m_t=173$\,GeV, such that the two-loop amplitudes
are only functions of two independent variables, the parton-level Mandelstam invariants
$\hat{s}$ and $\hat{t}$.  We have constructed a grid in these
variables, based on 5291 pre-computed phase-space points, together with an interpolation framework, such that an
external program can call the virtual two-loop amplitude at any phase space
point without having to do costly two-loop integrations.
We used the same setup for the grid as described in Ref.~\cite{Heinrich:2017kxx} and extended it in the following way:
We can write the squared matrix element as a polynomial of degree two in $\lambda$, 
\begin{align}
M_\lambda& \equiv |{\cal M}_\lambda|^2=A+B\,\lambda + C\,\lambda^2\;. \label{eq-amplambdadep}
\end{align}
Therefore it is sufficient to know the amplitude at three different values of $\lambda$ in order to reconstruct the full $\lambda$-dependence. 
Choosing $\lambda=-1,0,1$ we obtain
\begin{align}
A&=M_0\;,\; B=(M_1-M_{-1})/2\;,\; C=(M_1+M_{-1})/2-M_0\;.
\end{align}
In practice we used the representation 
\begin{align}
M_\lambda &=M_0\,(1-\lambda^2)+\frac{M_1}{2}\,(\lambda+\lambda^2) + \frac{M_{-1}}{2}\,(-\lambda+\lambda^2)\;
\end{align}
in order to get a more straightforward uncertainty estimate.

In fact, to any order in QCD,  we can separate the matrix element into a 
piece that depends only on the top quark Yukawa coupling $y_t$ (``box diagrams'') and a 
piece that depends on the Higgs boson trilinear self-coupling $\lambda$ (``triangle diagrams''):
\begin{align}
{\cal M} = y_t^2 {\cal M}_B + y_t \lambda {\cal M}_T.
\end{align}
The squared amplitude at each order can then be written as
\begin{align}
|{\cal M}|^2 = y_t^4 \left[ {\cal M}_B {\cal M}_B^* + \frac{\lambda}{y_t} ({\cal M}_B {\cal M}_T^* + {\cal M}_T {\cal M}_B^* ) +  \frac{\lambda^2}{y_t^2} {\cal M}_T {\cal M}_T^*  \right].\label{eq:yt}
\end{align}
The above parametrisation makes it clear that the dependence of the cross section on
both the Yukawa coupling and the Higgs boson self-coupling can be reconstructed
from only the 3 terms present in Eq.~(\ref{eq-amplambdadep}).
Of course this pattern changes once electroweak corrections, part of which have been calculated recently~\cite{Bizon:2018syu,Borowka:2018pxx}, are included. 

In order to allow for comparisons and cross checks, we implemented
both the $m_t\to\infty$ limit as well as the  amplitudes with full $m_t$-dependence at
NLO. This allows to run the code in four different modes by changing
the flag {\tt mtdep} in the \powhegbox{} run card. The possible
choices are the following:
\begin{description}
 \item[{\tt mtdep=0}:]{computation using basic HEFT: all amplitudes
   are computed in the $m_t\to\infty$ limit.}
\item[{\tt mtdep=1}:]{computation using Born-improved HEFT. In this
   approximation the NLO part is computed in the $m_t\to\infty$ limit
   and reweighted pointwise in the phase-space by the ratio of the LO matrix
   element with full mass dependence to  the LO matrix
   element in HEFT.}
 \item[{\tt mtdep=2}:]{computation in the approximation \ftapprox. In
   this approximation the matrix elements for the Born and the real
   radiation contributions are computed with full top quark mass dependence, whereas the virtual part is
   computed as in the Born-improved HEFT case. }
 \item[{\tt mtdep=3}:]{computation with full top quark mass dependence.}
\end{description}
Detailed instructions on how to run the code can be found in the
file {\tt manual-BOX-HH.pdf } in the folder {\tt ggHH/Docs} of the program.

When {\tt mtdep=3} is selected, the result of the virtual matrix element is based on a grid of pre-sampled phase-space points as described above. The phase-space points present in the grid are distributed such that they optimally sample the Standard Model (SM) Born matrix element. The same set of points is used regardless of the value of $\lambda$ selected. Due to the finite number of points present in the grid, there is an associated statistical uncertainty which amounts to $0.1\%$ on the total cross section at $14\ \TeV$ for $\lambda=\lambda_\mathrm{SM}$. However, for $\lambda \neq \lambda_\mathrm{SM}$ the virtual matrix element can differ significantly in shape from the SM prediction, as is apparent from examining the $\mhh$ and $\pth$ distributions for different values of the Higgs boson self coupling. The uncertainty associated with the use of the grid is therefore larger for non-SM values of $\lambda$. The uncertainty increases as $\lambda$ is decreased below the SM value reaching $0.6\%$ on the total cross section at $14\ \TeV$ for $\chhh = -1$. Increasing $\lambda$ above the SM value, we obtain an uncertainty of $3\%$ on the total cross section at $14\ \TeV$ for $\chhh = 3$ and  $\chhh = 5$. Furthermore, for differential distributions the total uncertainty is not distributed uniformly in each bin but instead increases when the shape of the matrix element most differs from the SM prediction. Focusing on the invariant mass distribution, amongst the values of the Higgs boson self-coupling considered here, the largest uncertainty is obtained for the smallest values of $\mhh$ and $\chhh=3$. The uncertainty reaches $6\%$ for the lowest bin when a $40\ \GeV$ bin width is used.

\section{Total and differential cross sections at non-SM trilinear couplings}
\label{sec:results}

The results were obtained using the
PDF4LHC15{\tt\_}nlo{\tt\_}30{\tt\_}pdfas~\cite{Butterworth:2015oua,CT14,MMHT14,NNPDF}
parton distribution functions interfaced to our code via
LHAPDF~\cite{Buckley:2014ana}, along with the corresponding value for
$\alpha_s$.  The masses of the Higgs boson and the top quark have been
fixed, as in the virtual amplitude, to $m_h=125$\,GeV, $m_t=173$\,GeV and their widths have been set to zero.
The top quark mass in renormalised in the on-shell scheme.
Jets are clustered with the
anti-$k_T$ algorithm~\cite{Cacciari:2008gp} as implemented in the
{\tt fastjet} package~\cite{Cacciari:2005hq, Cacciari:2011ma}, with jet
radius $R=0.4$ and a minimum transverse momentum 
$p_{T,\mathrm{min}}^{\rm{jet}}=20$\,GeV.  The scale uncertainties are
estimated by varying the factorisation/renormalisation scales
$\mu_{F}, \mu_{R}$. The scale variation bands 
represent scale variations around the central scale $\mu_0 =\mhh/2$, with
$\mu_{R} = \mu_{F}=c\,\mu_0$, where $c \in \{0.5,1,2\}$.
For the case $\lambda=\lambda_{\mathrm{SM}}$ we checked that the bands
obtained from these variations coincide with the bands resulting from
7-point scale variations. The PDF uncertainties have been studied in
\cite{deFlorian:2016spz} and found to be in general considerably smaller than the scale uncertainties.

\subsection{Total cross sections at different values of the trilinear coupling}

\begin{figure}[htb]
  \centering
    \includegraphics[width=0.65\textwidth]{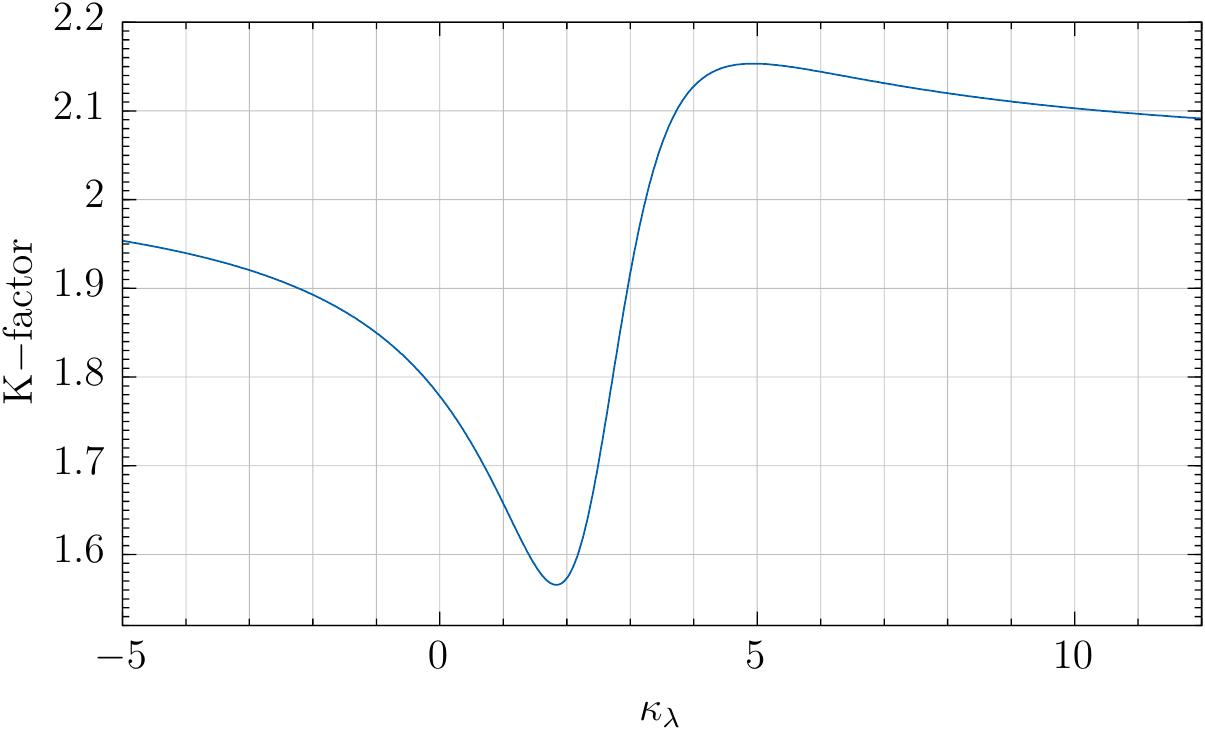}
%  \includegraphics[width=\textwidth]{plots/}
%    \caption{\label{fig:lambda_large_27}}
\caption{Variation of the NLO K-factor with the trilinear coupling at $\sqrt{s}=14$\,TeV.}
\label{fig:Kfacvariation}
\end{figure}

In Table \ref{tab:sigmatot} we list total cross sections at 13, 14 and 27\,TeV for various values of the trilinear Higgs coupling $\lambda$. 
\begin{table}[htb]
\begin{center}
\begin{tabular}{| c | c | c |c|c|}
%\Xhline{2\arrayrulewidth}
\hline
&&&&\\
$\lambda_{\mathrm{BSM}}/\lambda_{\mathrm{SM}}$ & $\sigma_{\rm{NLO}}@13 \mathrm{TeV}$\,[fb]& $\sigma_{\rm{NLO}}@14 \mathrm{TeV}$\,[fb] & $\sigma_{\rm{NLO}}@27 \mathrm{TeV}$\,[fb] &K-factor@14TeV\\
&&&&\\
\hline
-1& 116.71$^{+16.4\%}_{-14.3\%}$  & 136.91$^{+16.4\%}_{-13.9\%}$& 504.9$^{+14.1\%}_{-11.8\%}$ & 1.86 \\
\hline
0& 62.51$^{+15.8\%}_{-13.7\%}$ & 73.64$^{+15.4\%}_{-13.4\%}$& 275.29$^{+13.2\%}_{-11.3\%}$& 1.79  \\
\hline 
1& 27.84$^{+11.6\%}_{-12.9\%}$ & 32.88$^{+13.5\%}_{-12.5\%}$&127.7$^{+11.5\%}_{-10.4\%}$ &1.66\\
\hline
2 & 12.42$^{+13.1\%}_{-12.0\%}$ & 14.75$^{+12.0\%}_{-11.8\%}$ &  59.10$^{+10.2\%}_{-9.7\%}$ & 1.56 \\
\hline
2.4& 11.65$^{+13.9\%}_{-12.7\%}$ & 13.79$^{+13.5\%}_{-12.5\%}$& 53.67$^{+11.4\%}_{-10.3\%}$ & 1.65 \\
\hline
3& 16.28$^{+16.2\%}_{-15.3\%}$ & 19.07$^{+17.1\%}_{-14.1\%}$ & 69.84$^{+14.6\%}_{-12.1\%}$ & 1.90 \\
\hline 
5& 81.74$^{+20.0\%}_{-15.6\%}$  & 95.22$^{+19.7\%}_{-11.5\%}$& 330.61$^{+17.4\%}_{-13.6\%}$ & 2.14 \\
\hline 
\end{tabular}
\end{center}
\caption{Total cross sections for Higgs boson pair production at full NLO QCD. The given uncertainties are scale uncertainties. 
\label{tab:sigmatot}}
\end{table}
Table~\ref{tab:sigmatot} also shows that the K-factors vary substantially as functions of the trilinear coupling.
This fact is illustrated in Fig.~\ref{fig:Kfacvariation}, showing that the K-factor takes values between 1.56 and 2.15
if the trilinear coupling is varied between $-5\leq \chhh\leq 12$.

\subsection{Differential cross sections}

\begin{figure}[htb]
 \begin{subfigure}[t]{0.495\textwidth}
\includegraphics[width=\textwidth]{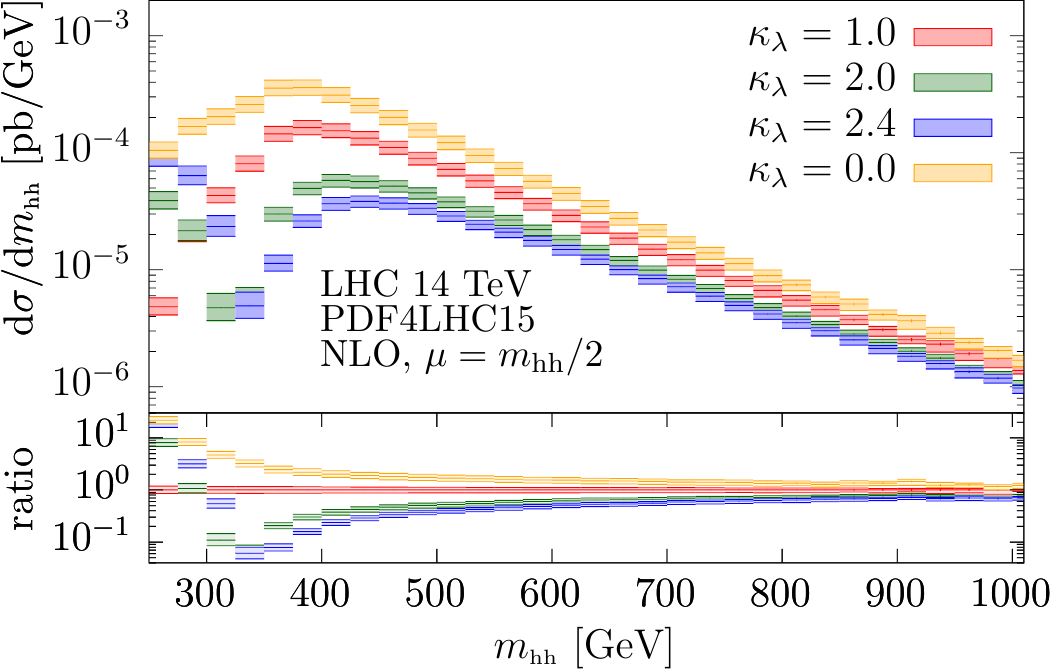}
%    \vspace{\TwoFigBottom em}
\caption{\label{fig:lambda_small}}
\end{subfigure}
\hfill
\begin{subfigure}[t]{0.495\textwidth}
    \includegraphics[width=\textwidth]{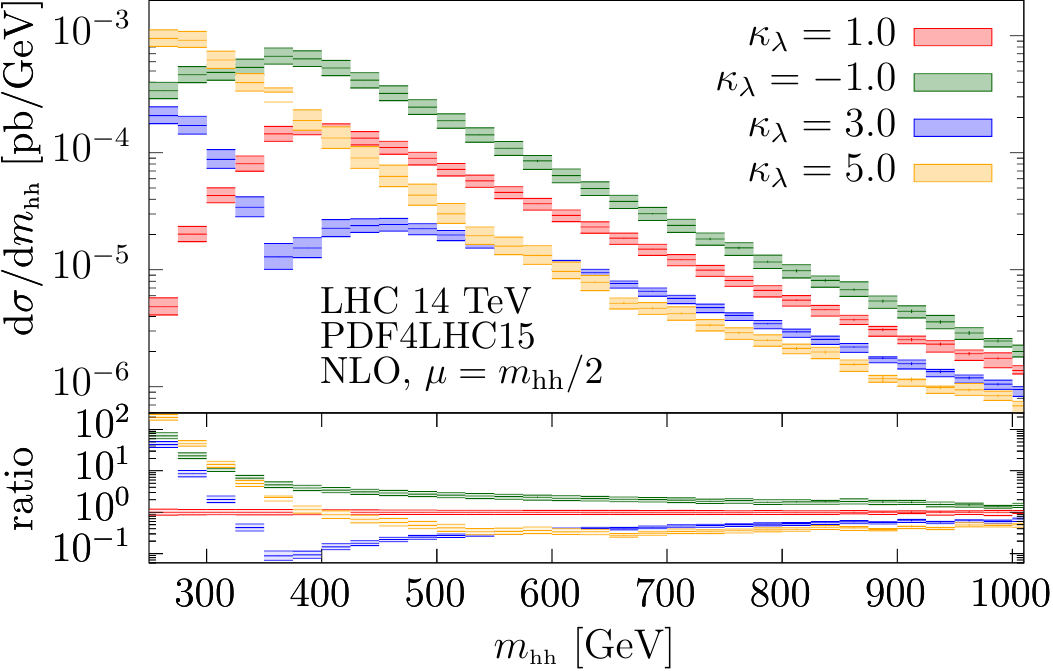}
\caption{\label{fig:lambda_large}}
\end{subfigure}
\caption{Higgs boson pair invariant mass distributions for various
  values of $\chhh$  at $\sqrt{s}=14$\,TeV. The uncertainty bands are from
  scale variations as described in the text.}
\label{fig:lambdavar14TeV}
\end{figure}

In Fig.~\ref{fig:lambdavar14TeV} we show the $\mhh$ distribution for various values of $\chhh=\lambda_{\mathrm{BSM}}/\lambda_{\mathrm{SM}}$. 
The ratio plots show the ratio to the result with $\lambda_{\mathrm{SM}}$. 
A characteristic dip develops in the $\mhh$ distribution around $\chhh=2.4$, which is the value of maximal destructive interference between diagrams containing the trilinear coupling (triangle-type contributions) and ``background" diagrams (box-type contributions).
Therefore we provide results for a denser spacing of $\chhh$ values around this point.

\begin{figure}[htb]
 \begin{subfigure}[t]{0.495\textwidth}
\includegraphics[width=\textwidth]{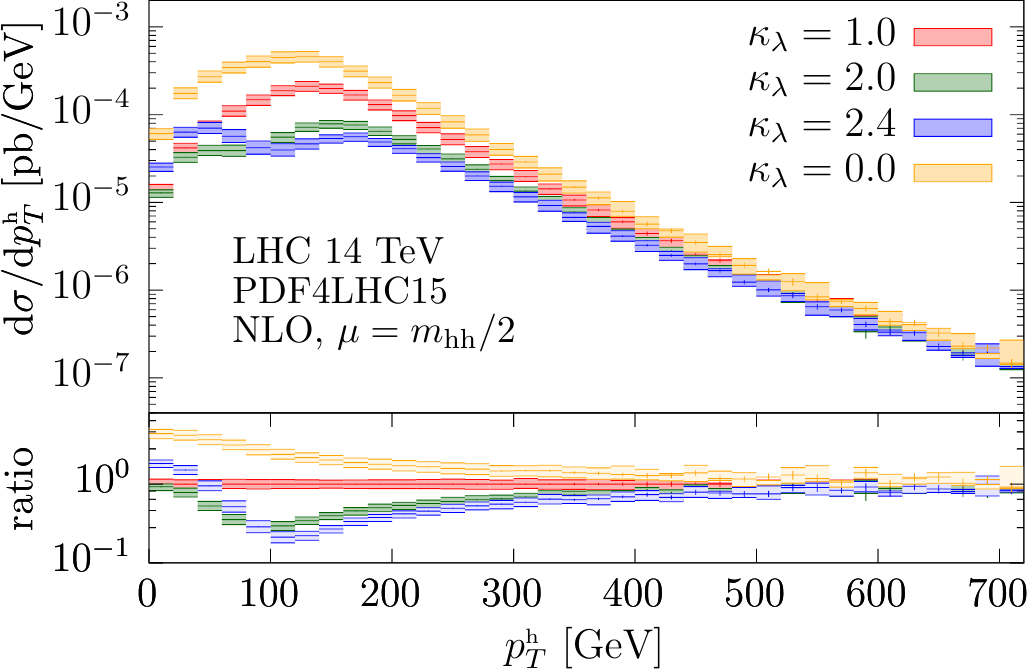}
\caption{\label{fig:lambda_small_pTH}}
\end{subfigure}
\hfill
\begin{subfigure}[t]{0.495\textwidth}
    \includegraphics[width=\textwidth]{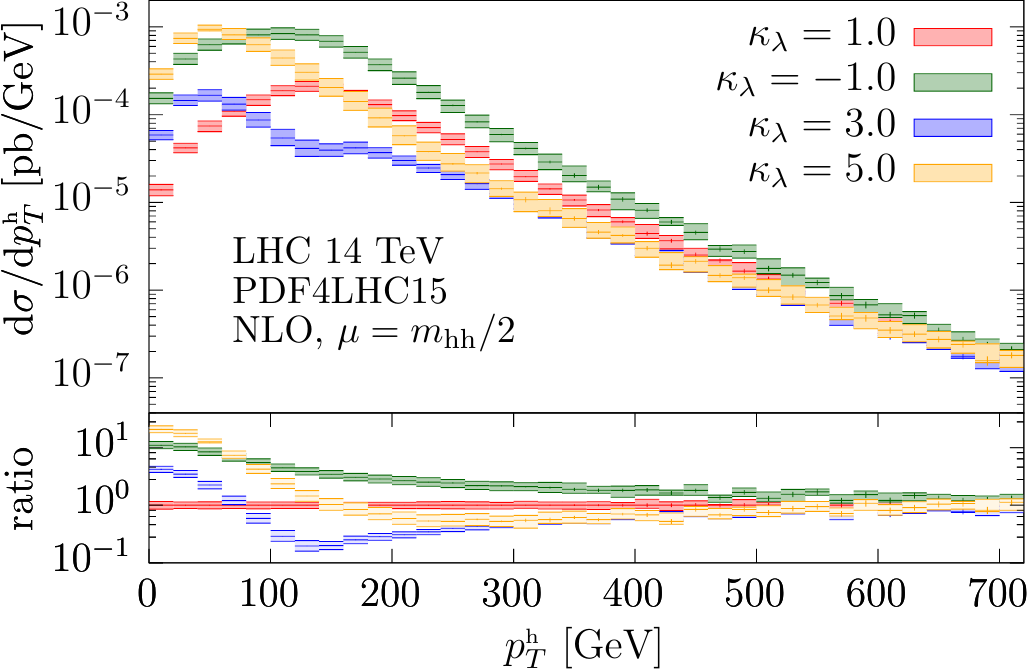}
\caption{\label{fig:lambda_large_pTH}}
\end{subfigure}
\caption{Higgs boson transverse momentum distributions for various values of $\chhh$  at $\sqrt{s}=14$\,TeV.
\label{fig:lambdavar14TeV_pTH}}
\end{figure}
In Fig.~\ref{fig:lambdavar14TeV_pTH} we show the transverse momentum
distributions $p_T^h$ of one (any) Higgs boson for different $\chhh$
values. The dip for $\chhh\sim 2.4$ is still present, however much
less pronounced than in the $\mhh$ distribution.

\begin{figure}[htb]
 \begin{subfigure}[t]{0.495\textwidth}
\includegraphics[width=\textwidth]{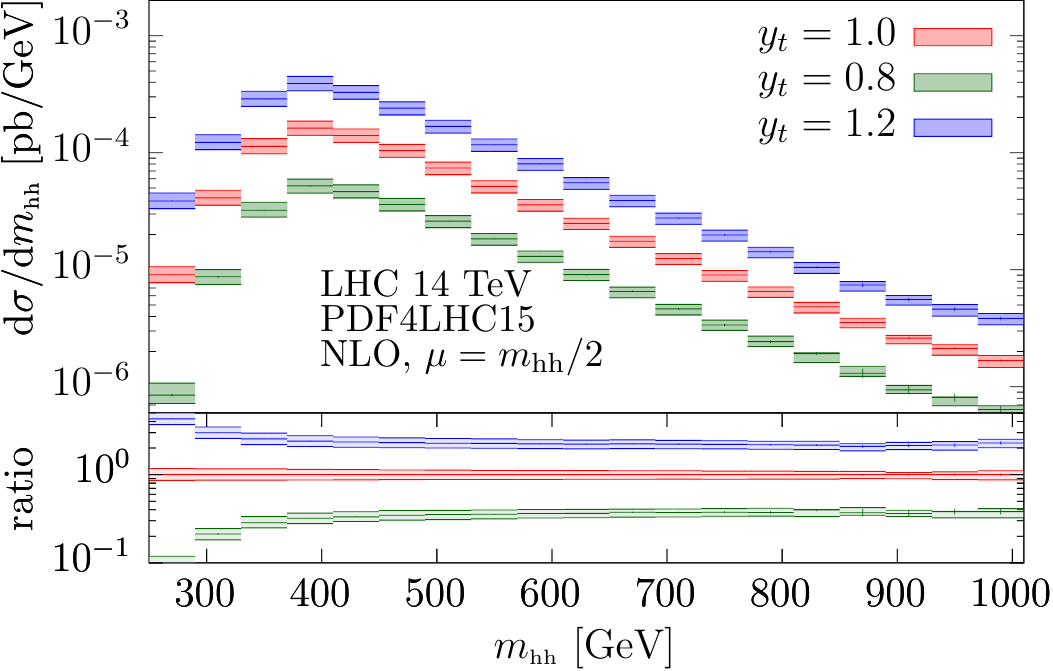}
\caption{\label{fig:ytvar_mhh}}
\end{subfigure}
\hfill
\begin{subfigure}[t]{0.495\textwidth}
\includegraphics[width=\textwidth]{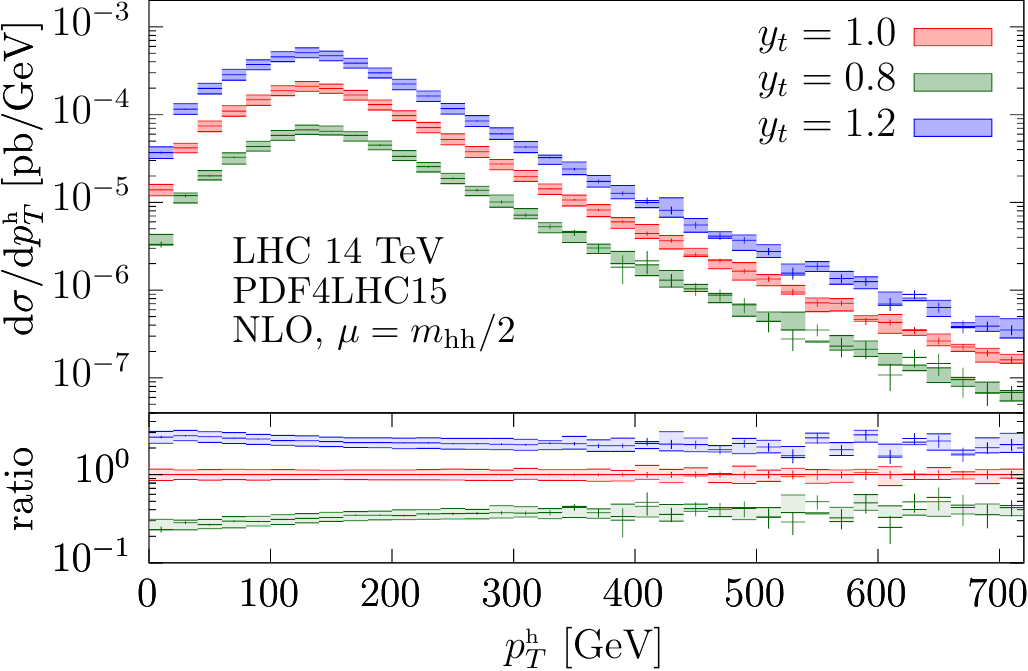}
\caption{\label{fig:ytvar_pth}}
\end{subfigure}
\caption{Higgs boson pair invariant mass distributions, and distributions of the transverse momentum of one (any) Higgs boson for non-SM values of the top quark Yukawa coupling $y_t$  at $\sqrt{s}=14$\,TeV, including scale uncertainties.
\label{fig:ytvar}}
\end{figure}

Fig.~\ref{fig:ytvar} demonstrates the effect of variations of the top quark Yukawa coupling $y_t$ on the $\mhh$ and $\pth$ distributions, where $\chhh$ is fixed to the SM value.
Using eq.~(\ref{eq:yt}), it is apparent that $y_t$ variations can be obtained from appropriate $\chhh$ variations with the same code.
For example, $\sigma(y_t=1.2,\chhh=1)=(1.2)^4\,\sigma(y_t=1,\chhh=1/1.2)$.

\subsection{Discussion of parton shower related uncertainties} 
 
In this section we show distributions for NLO results matched to a
parton shower, focusing mostly on the transverse momentum of the Higgs
boson pair. For this distribution 
NLO is the first non-trivial order, and therefore it is particularly
sensitive to differences in the treatment of radiation by the
parton shower.
We compare the \pythia~\cite{Sjostrand:2014zea} and \herwig~\cite{Bellm:2017bvx} parton showers, 
applied directly to the \powheg{} Les Houches events (LHE).
In the \herw case, we also compare the default shower (the angular-ordered $\tilde{q}$-shower) 
with the dipole shower.
In addition, we assess the uncertainties stemming from the matching and show results where the 
\herw shower scale parameter {\tt HardScale} is varied. For all shower algorithms considered, the default 
tune of the corresponding version is used. Multiple-parton interactions (MPI) and hadronisation 
are switched off. The \hdamp{} parameter in \powheg{} is set to $\hdamp=250$\,GeV.

\begin{figure}[htb]
 \begin{subfigure}[t]{0.495\textwidth}
\includegraphics[width=\textwidth]{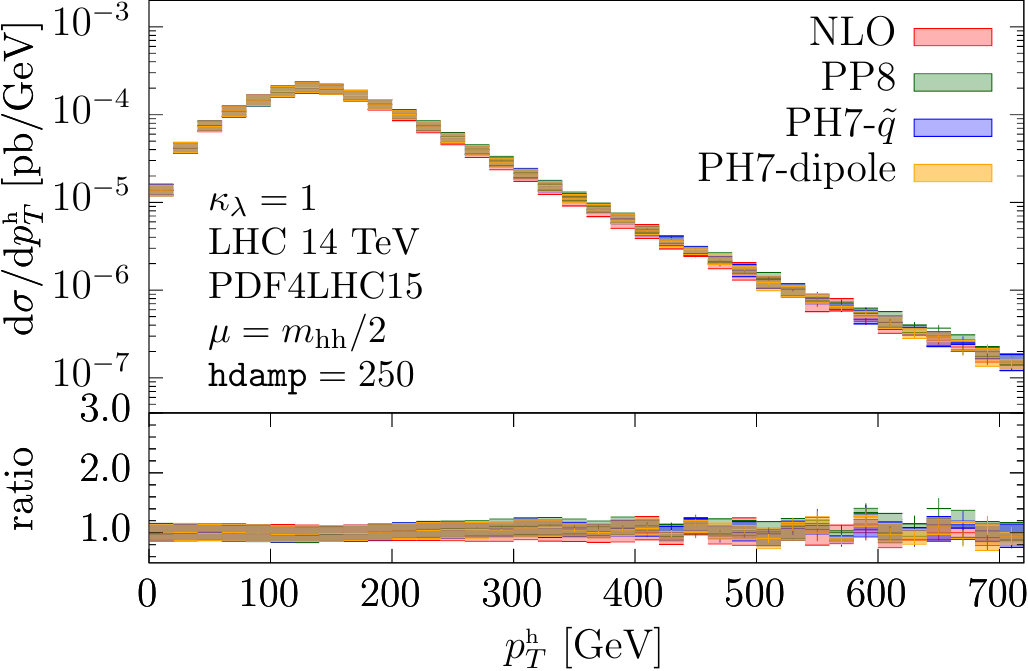}
 \caption{\label{fig:lambda_pTh}}
\end{subfigure}
\hfill
\begin{subfigure}[t]{0.495\textwidth}
    \includegraphics[width=\textwidth]{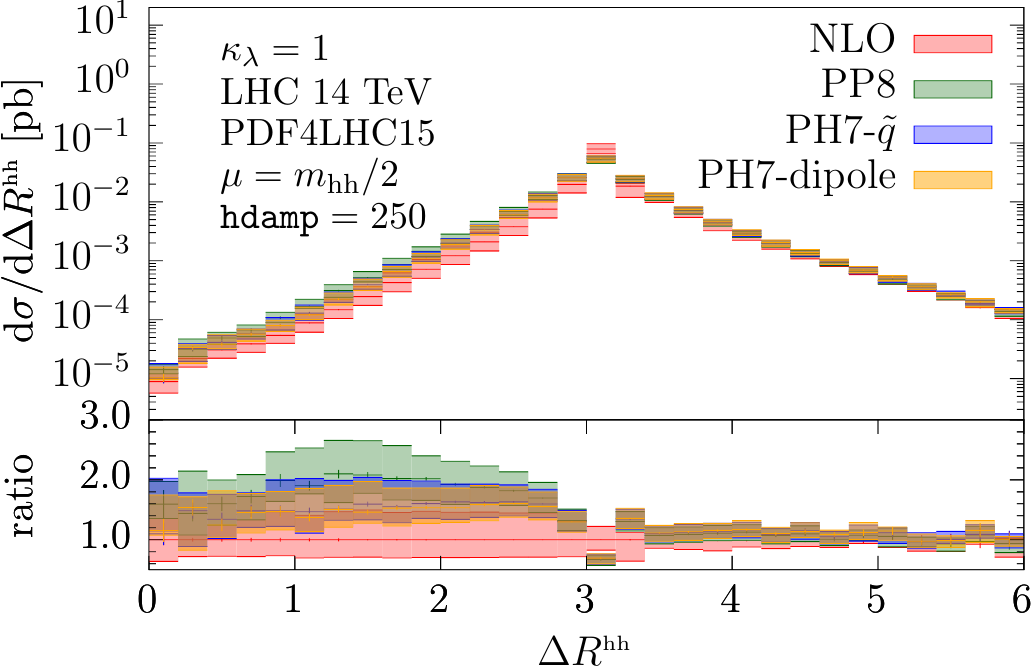}
\caption{\label{fig:lambda_dRHH}}
\end{subfigure}
\caption{The transverse momentum of one (any) Higgs boson and the $R$-separation between 
the two Higgs bosons are shown for the fixed-order NLO calculation and three shower setups, in 
the $\chhh=1$ case.
\label{fig:lambdavar14TeV_pTH_dRHH_showers}}
\end{figure}

In general, observables that are inclusive in the additional radiation, like the 
transverse momentum of one (any) Higgs boson, $p_T^h$, show little sensitivity to the details of 
the parton showering, as can be seen from Fig.~\ref{fig:lambda_pTh}, showing the fixed-order NLO prediction, 
as well as  the \pythia (PP8) and both \herwig showers (angular-ordered PH7-$\tilde{q}$, 
and PH7-dipole). In contrast, Fig.~\ref{fig:lambda_dRHH} 
displays the distribution of the distance $\Delta
R^{\mathrm{hh}}=\sqrt{(\eta_1-\eta_2)^2+(\Phi_1-\Phi_2)^2}$ between the two Higgs 
bosons. There, the Sudakov exponent and the parton shower effectively resum the fixed-order 
prediction in the region where the two Higgs bosons are close to a back-to-back configuration, 
and the parton shower increases the fixed-order real radiation contribution in the region $\Delta R^{\mathrm{hh}}<\pi$.

\begin{figure}[htb]
 \begin{subfigure}[t]{0.495\textwidth}
\includegraphics[width=\textwidth]{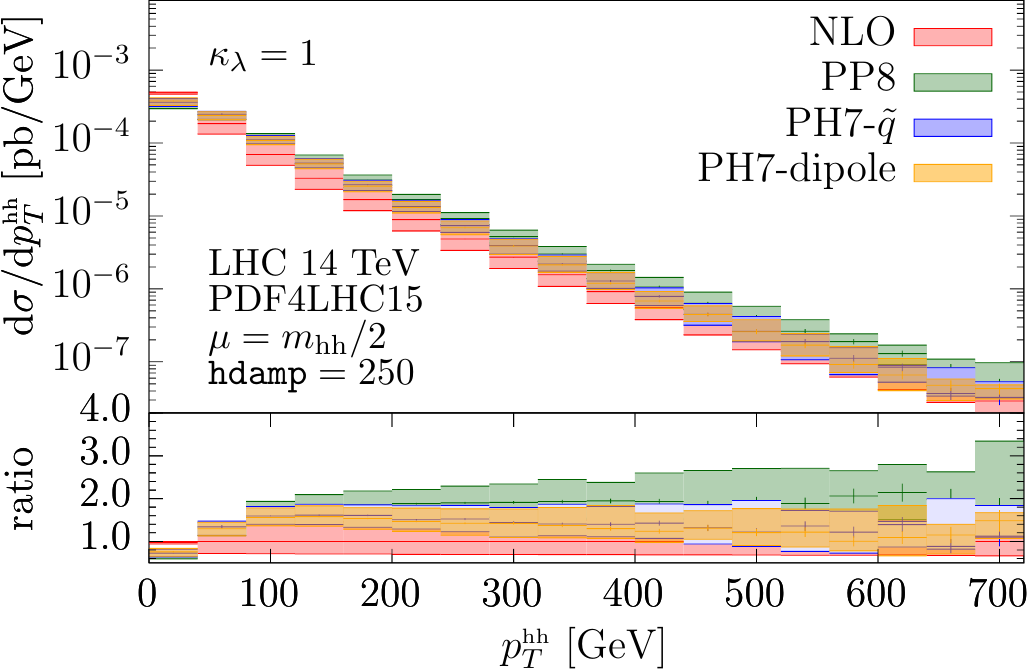}
 \caption{\label{fig:lambda_ptHH_cHHH1}}
\end{subfigure}
\hfill
\begin{subfigure}[t]{0.495\textwidth}
    \includegraphics[width=\textwidth]{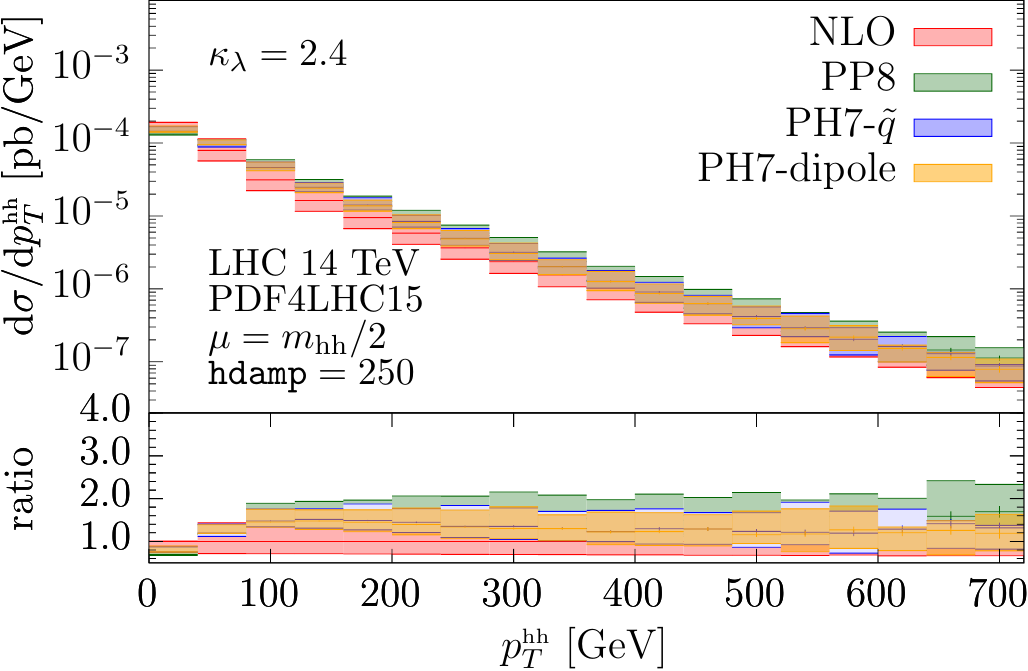}
\caption{\label{fig:lambda_ptHH_cHHH2.4}}
\end{subfigure}
\caption{Transverse momentum of the Higgs boson pair for the fixed-order NLO calculation 
and all three shower setups at 14\,TeV for (a) $\chhh=1$, (b) $\chhh=2.4$.
\label{fig:lambdavar14TeV_pTHH_showers}}
\end{figure}

In Figs.~\ref{fig:lambda_ptHH_cHHH1} and \ref{fig:lambda_ptHH_cHHH2.4}, the transverse 
momentum $\pthh$ of the Higgs boson pair system is shown for the fixed-order and parton-showered predictions, at $\chhh=1$ and $\chhh=2.4$. In all cases, the \pyth and \herw showers 
agree very well in the small-$\pthh$ range, but start to deviate already at $\pthh \sim 100$\,GeV. 
While both \herw showers give very similar results and reproduce the fixed-order calculation at high-$\pthh$, the 
\pyth shower produces much harder additional radiation and the ratio to the fixed-order result plateaus at $\sim 2.0$ over the remaining range. 
%A discussion of the surprisingly hard tail of the $\pthh$ distribution with \powheg+\pythia can be found in
%Refs.~\cite{Jones:2017giv,Bendavid:2018nar}, where the results are also
%compared to two different showers within {\sc Sherpa}~\cite{Gleisberg:2008ta,Schumann:2007mg,Hoche:2015sya} as well
%as with {\sc MG5\_aMC@NLO}~\cite{Alwall:2014hca,Hirschi:2015iia}.
We should mention that rather large differences between \pythia and
\herwig showers matched to \powheg{} also have been found studying top
quark pair production~\cite{Ravasio:2018lzi}. The origin of the large NLO parton shower matching uncertainties affecting certain observables in Higgs boson pair production have previously been studied in literature~\cite{Jones:2017giv}. For the SM result, the excess at large $\pthh$ produced when using \powheg{} with \pythia was found to be due to additional hard sub-leading jets generated purely by the shower~\cite{Bendavid:2018nar}.

\begin{figure}[htb]
 \begin{subfigure}[t]{0.495\textwidth}
\includegraphics[width=\textwidth]{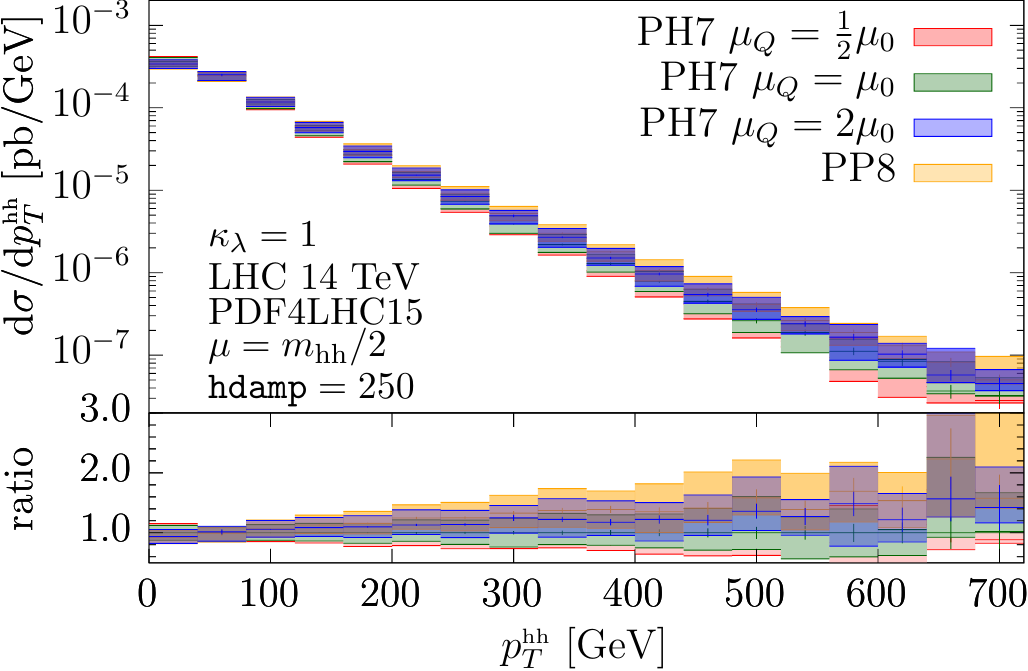}
 \caption{\label{fig:lambda_ptHH_muQvar}}
\end{subfigure}
\hfill
\begin{subfigure}[t]{0.495\textwidth}
    \includegraphics[width=\textwidth]{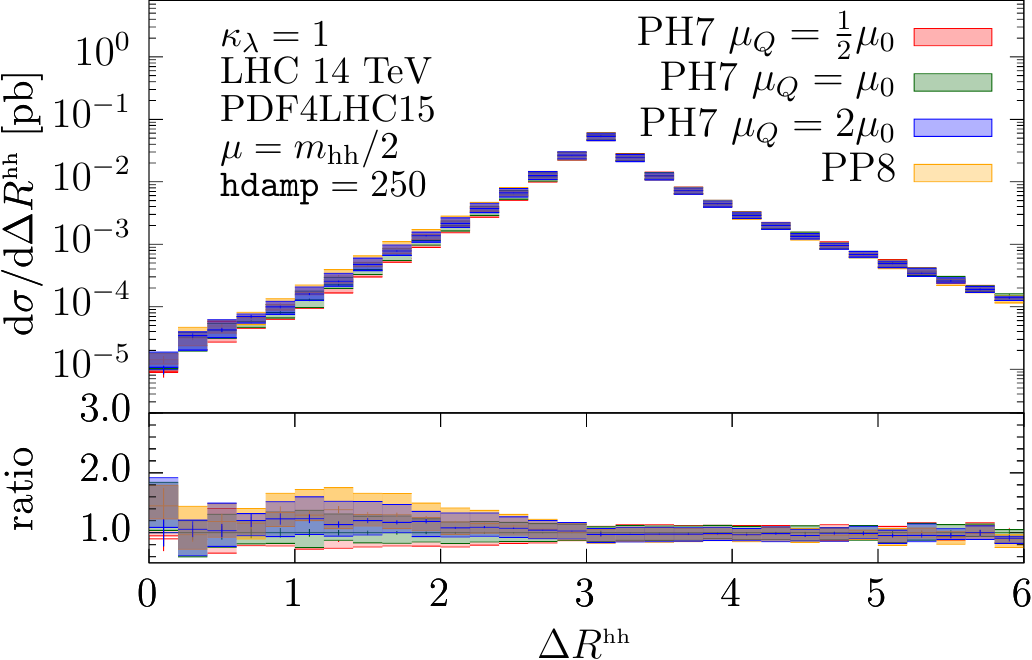}
\caption{\label{fig:lambda_dRHH_muQvar}}\end{subfigure}
\caption{Higgs boson pair transverse momentum and $R$-separation for variations of the \herw $\tilde{q}$-shower hard scale.
\label{fig:lambdavar14TeV_muQvar}}
\end{figure}

With the \herw default shower, systematic uncertainties can be estimated by 
varying the maximal transverse momentum allowed for shower emissions,
by changing the so-called hard scale $\mu_Q$. We apply a factor $c_Q=\{0.5,\,2.0\}$ on the central hard shower scale, separately 
for all variations of the factorisation/renormalisation scales
$\mu_{R,F}$.
%, which is the standard use in the \herw setup. 
Fig.~\ref{fig:lambdavar14TeV_muQvar} shows the $\pthh$ and 
$\Delta R^{\mathrm{hh}}$ distributions as examples of the SM case, $\chhh=1$, and underlines 
their sensitivity to changes in the shower hard scale. Quantitatively, the hard scale variations 
inflate the sole factorisation/renormalisation scale uncertainties by a factor of two in 
the regions where the \herwig and \pythia showers were in disagreement (see Figs.~\ref{fig:lambda_dRHH} 
and \ref{fig:lambdavar14TeV_pTHH_showers}). If the envelope of all scale variations, including 
the hard shower scale, was to be taken as a theoretical systematic uncertainty, the resulting 
uncertainty would be of the order of 50\% in these bins. It would be enlightening to further study 
parton shower (and non-perturbative) effects, in the particular context of Higgs boson pair 
production at NLO, as well as for loop-induced colour singlet production in general, and try to reduce discrepancies among the different algorithms.

\section{Conclusions}
\label{sec:conclusions}

We have presented results for Higgs boson pair production in gluon
fusion at full NLO QCD for non-standard values of the trilinear Higgs
boson coupling $\lambda$. We have also shown how results with a modified 
top quark Yukawa coupling can be produced with the same code.

We have demonstrated that the dependence of both the total and the
differential K-factors on the value of $\lambda$ is stronger than the $m_t\to\infty$ limit may suggest.
The total cross section is a quadratic
polynomial in $\lambda$, with a minimum around $\chhh\approx 2.4$,
which is present both at LO and NLO with full top quark mass
dependence, stemming from destructive interference of diagrams with
and without a trilinear Higgs coupling. 
The $\mhh$ distribution shows a dip around this minimum, which is to
lesser extent also visible in the transverse momentum distribution of
one of the Higgs bosons. 
We have assumed in our study that modifications of the Higgs couplings
to other particles are small and can be increasingly well constrained by
other processes. Nonetheless, it should be kept in mind that a dip in
the $\mhh$ distribution could also originate from other effective
couplings, for example an effective $t\bar{t}HH$ coupling, while
$\chhh= 1$~\cite{Buchalla:2018yce}.

We have also combined our NLO QCD results with the \pythia and \herwig
parton showers. In the \herwig case we employed both the default 
shower (the angular-ordered $\tilde{q}$-shower) and the dipole shower.
We observed that for distributions particularly sensitive to the 
additional radiation, the parton showers exhibit a somewhat different 
behaviour. While both \herwig showers generate comparable results and 
perform as expected in the NLO regime, the \pythia shower produces harder 
radiation, for example in the tail of the $\pthh$
distribution. Varying the shower hard scale in \herwig on top of
$\mu_R,\mu_F$ variations leads to
uncertainty bands which approximately cover these 
differences. However, the parton shower uncertainties 
can then become sizeable and even surpass the fixed-order scale uncertainties.

The \powheg{} version of the code for Higgs boson pair production
including the possibility to vary the trilinear coupling 
and the top quark Yukawa coupling 
is publicly available in the \powhegbox{\tt-V2} package at the website
{\tt http://powhegbox.mib.infn.it}, in the
{\tt User-Processes-V2/ggHH/} directory.

\section*{Acknowledgements}
We would like to thank Gerhard Buchalla, Alejandro Celis, Matteo
Capozi, Stephan Jahn and Emmanuele Re for helpful discussions.
This research was supported in part by the COST Action CA16201 `Particleface' of the European Union, 
and by the Swiss National Science Foundation (SNF) under grant number 200020-175595.
We gratefully acknowledge resources provided by the Max Planck Computing and Data Facility (MPCDF).

%\clearpage

%%%%%%%%%%%%%%%%%%%%%%%%%%%%%%%%%%%%%%%%%%%%%%%%%%%%%%%%%%%%%% 
\bibliographystyle{JHEP}
 
\providecommand{\href}[2]{#2}\begingroup\raggedright\endgroup

\end{document}